\documentclass[aps,prl,twocolumn,groupedaddress,showpacs]{revtex4}

\usepackage{amssymb,amsmath}

\begin{document}

\title{
Comment on ``Observation of Collective Modes of Ultracold Plasmas''
}

\author{Yurii V. Dumin}

\email[Electronic address: ]{dumin@yahoo.com}

\affiliation{
Theoretical Department, IZMIRAN, Russian Academy of Sciences,
Troitsk, Moscow reg., 142190 Russia
}

\date{June 28, 2009}

\pacs{32.80.Rm, 42.50.Ct, 34.80.Lx, 52.27.Gr}
%

\maketitle

In the work~\cite{fle06} Fletcher \textit{et al.}\ reported on
the new interesting phenomenon in ultracold plasmas---the observation of
subharmonics of the electron emission from the expanding plasma clouds
under their irradiation by the external radiowaves
(in addition to the basic plasmon harmonic detected in the earlier works).
Unfortunately, a theoretical interpretation of the additional harmonics
remained doubtful.

The authors of the experiment assumed that the subharmonics were
the so-called Tonks--Dattner reso\-nances---the standing waves of
the electron plasma oscillations arising due to the thermal term in
the Bohm--Gross dispersion relation. Such interpretation, however,
encounters a number of obstacles. One of them was mentioned already
in the original paper~\cite{fle06}: this is the lack of the adequate
boundary conditions for the formation of standing waves in
the unbounded gaseous cloud.

The second problem, which may be even more important, is the
temperature dependence of the observed subharmonics.
If they are produced by a thermal term in the dispersion relation,
the pattern of subpeaks in Fig.~1~\cite{fle06} should be more pronounced
with the increase in the electron temperature and, therefore,
decrease in the neutrality of the plasma cloud (due to escape of
the most energetic electrons). This seems to be in conflict with
the experimental finding that ``the largest number of peaks are
found at the highest densities and neutralities''.

The aim of the present comment is to mention that a more promising
interpretation of the observed phenomenon might be based on
the ionization of secondary Rydberg atoms, formed during expansion
and cooling of the plasma cloud. This idea is well supported,
particularly, by a close similarity between the subharmonics of
electron emission presented in Fig.~1 of paper~\cite{fle06} and
in Fig.~3 of the work by Maeda and Gallagher~\cite{mae04}, where
no plasma effects were involved at all.

In the last-cited experiment, the atoms were excited to the specified
Rydberg states, and it was found that the efficiency of their
subsequent ionization by microwave (MW) irradiation shows a number of
peaks in the interval of the scaled frequency~$ {\Omega} \equiv
{\omega} / {\omega}_{\rm K} \approx 0.5{\div}2 $, where
$ \omega $~is the MW frequency, and $ {\omega}_{\rm K} $~is
the Keplerian frequency of the atom.
On the other hand, in the experiment with plasma clouds~\cite{fle06}
there should be formation of the secondary Rydberg atoms due to
three-body recombination during the cloud expansion and cooling.
The typical radii of such atoms are initially just about
the average electron--ion separation in plasma
(Wigner--Seitz radius)~$ r_{\rm WS} \approx (2N)^{-1/3} $,
where $ N $~is the concentration of charged particles; and
their Keplerian frequencies should be
$ {\omega}_{\rm K} \equiv e / \big( m_e^{1/2} r_{\rm WS}^{3/2} \big)
\approx \sqrt{2} \, e \, m_e^{-1/2} N^{1/2} $.
This quantity almost coincides with a plasmon frequency for
the spherical cloud, which is the standard Langmuir
frequency~$ {\omega}_{\rm L} \equiv
2 \sqrt{\pi} \, e \, m_e^{-1/2} N^{1/2} $
reduced by the factor~$ 2{\div}3 $, depending on the particular
density distribution inside the cloud.

Therefore, the pattern of subharmonics observed in the ultracold
plasma experiment~\cite{fle06} might be a combination of
the basic plasmon oscillation (which should definitely exist)
and a number of peaks in the ionization efficiency of
the secondary Rydberg atoms. The peaks occur at the various
values of~$ \omega / {\omega}_{\rm K} $, when concentration~$ N $
changes in the course of the plasma cloud expansion.
Such interpretation avoids the problem of adequate boundary
conditions for the formation of standing waves and gives a correct
dependence on the plasma parameters (the secondary Rydberg atoms
should be formed more efficiently at the lower electron temperature
and, consequently, the higher plasma neutrality).

It is important to mention also that the oscillatory structure
of the ionization efficiency in the study by Maeda and
Gallagher was especially pronounced when the ionization potential
was artificially decreased (\textit{e.g.} curves for the threshold
principal quantum numbers $ n_c = 145 $ and 120 in
Fig.~3~\cite{mae04}). And essentially the same situation took
place in the experiment by Fletcher \textit{et al.}~\cite{fle06},
where the ionization threshold was decreased just due to
the finite separation between the plasma particles.

\end{document}